# The Einstein Postulates: 1905-2005

# A Critical Review of the Evidence

Reginald T. Cahill
School of Chemistry, Physics, Earth Sciences, Flinders University, GPO Box 2100, Adelaide 5001, Australia

The Einstein postulates assert an invariance of the propagation speed of light in vacuum for any observer, and which amounts to a presumed absence of any preferred frame. The postulates appear to be directly linked to relativistic effects which emerge from Einstein's *Special Theory of Relativity,* which is based upon the concept of a flat spacetime ontology, and which then lead to the *General Theory of Relativity* with its curved spacetime model for gravity. While the relativistic effects are well established experimentally it is now known that numerous experiments, beginning with the Michelson-Morley experiment of 1887, have always shown that the postulates themselves are false, namely that there is a detectable local preferred frame of reference. This critique briefly reviews the experimental evidence regarding the failure of the postulates, and the implications for our understanding of fundamental physics, and in particular for our understanding of gravity. A new theory of gravity is seen to be necessary, and this results in an explanation of the `dark matter' effect entailing the discovery that the fine structure constant is a 2nd gravitational constant.

## Introduction

It is one hundred years since Einstein formulated his postulates for the invariant property of light, namely that the speed of light is always c (≈300,000 km/s) for any uniformly moving observer, which is equivalent to the assertion that there is no preferred frame, that there is no detectable *space*, that a three-dimensional *space* has no physical existence.

> **Einstein postulates:**
> (1) The laws of physics have the same form in all inertial reference frames.
> (2) Light propagates through empty space with a definite speed c independent of the speed of the observer (or source).
> (3) In the limit of low speeds the gravity formalism should agree with Newtonian gravity.

The putative successes of the postulates lead to the almost universal acceptance of the Einstein *Special Theory of Relativity*, which is based upon the concept of a flat spacetime ontology that replaces the older separate concepts of space and time, and then to the *General Theory of Relativity* with its curved spacetime model for gravity. While the relativistic effects are well established experimentally it is now belatedly understood, in 2002 [4,10], that numerous experiments, beginning with the Michelson-Morley experiment [1] of 1887, have always shown that postulates (1) and (2) (excepting the 2nd part) are false, namely that there is a detectable local frame of reference or `space', and that the solar system has a large observed galactic velocity of some 420±30km/s in the direction (RA=5.2hr, Dec= -67deg) through this space [2,3,5,8,10]. This is different from the speed of 369km/s in the direction (RA=11.20hr, Dec= -7.22deg) extracted from the Cosmic Microwave Background (CMB) anisotropy, and which describes a motion relative to the distant universe, but not relative to the local space. This critique briefly reviews the experimental evidence regarding the failure of the postulates, and the implications for our understanding of fundamental physics, and in particular for our understanding of gravity. A new theory of gravity is seen to be necessary, and this results in an explanation of the `dark matter' effect, entailing the discovery that the fine structure constant is a 2nd gravitational constant [1-4]. This theory is a part of the information-theoretic modelling of reality known as *Process Physics* [2-4,9], which premises a non-geometric process model of time, as distinct from the current *non-Process Physics,* which is characterised by a geometrical model of time.

## Over 100 years of Detecting Absolute Motion

The whole business of detecting absolute motion (motion relative to space itself) and so a preferred local frame of reference, came undone from the very beginning.

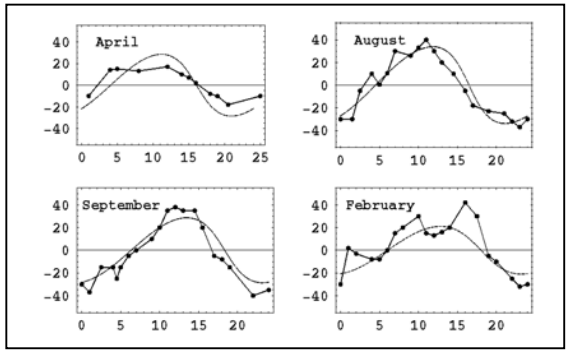

Fig.1 Azimuth $\psi$ (deg), measured from south, from the Miller [5] data, plotted against local sidereal time. Plots cross the local meridian at approx. 5hr and 17hr. The monthly changes arise from the orbital motion of the earth about the sun. Miller used that effect to determine the value of k in (1), and now in agreement with the refractive index theory, see [2,3,4]. Curves from theory [2,4].

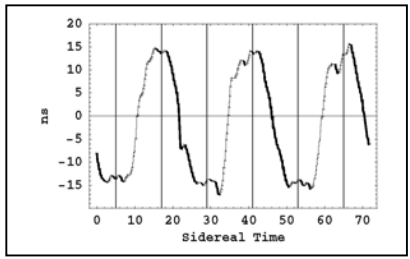

Fig.2 DeWitte 1991 RF travel time variations, in ns, in a 1.5km NS coaxial cable, measured with atomic clocks, over three days and plotted against local sidereal time, showing that at approximately 5hr and 17hr the effect is largest. This remarkable agreement with the Miller interferometer experiment shows that the detection of absolute motion is one of the great suppressed discoveries in physics. At least five other interferometer or coaxial cable experiments are consistent with these observations [2-4].

The Michelson and Morley air-mode interferometer fringe shift data revealed a speed of some 8km/s [4,10] when analysed using the prevailing Newtonian theory, which has $k^2 = n^3 \approx 1$ in (1), where $\Delta t$ is the difference between the light travel times for the two arms, effective length $L$, within the interferometer, and $n$ is the refractive index of the gas present.

$$\Delta t = k^2 \frac{L v_P^2}{c^2} \cos(2(\theta - \psi)) \tag{1}$$

However including the Fitzgerald-Lorentz dynamical contraction effect as well as the effect of the gas present we find that $k^2 = n(n^2 - 1)$, giving $k^2 = 0.00058$ for air, which explains why the observed fringe shifts were so small. In the Einstein theory $k = 0$; absolute motion is to be undetectable in principle. In (1) $\theta$ is the azimuth of one arm relative to the local meridian, with $\psi$ the azimuth of the projected absolute motion velocity $v_P$. Fig.1 shows $\psi$ from the 1925/1926 Miller [5] interferometer data for four different months of the year, from which the RA = 5.2hr is readily apparent. The orbital motion of the earth about the sun slightly affects the RA in each month, and Miller used this effect do determine the value of $k$, but the new theory of gravity required a reanalyse of his data [2-4]. Two interferometer experiments used helium, enabling the refractive index effect to be confirmed. Fig.2 shows the coaxial cable travel times measured by DeWitte in 1991, which also show the same RA [2-4]. That these very different experiments show the same speed and RA of absolute motion is one of the most startling but suppressed discoveries of the twentieth century.

So Postulates (1) and (2) are in disagreement with the experimental data. In all some seven experiments have detected this absolute motion. Modern interferometer experiments use vacuum with $n = 1$, and then from (1) $k = 0$, predicting no fringe shifts. In analyzing the data this is misinterpreted to imply the absence of absolute motion. As discussed in [2-4] it is absolute motion which causes the dynamical effects of length contractions, time dilations and other relativistic effects, in accord with Lorentzian interpretation of relativity.

## Gravity as Inhomogeneous and Time-Dependent Spatial Flows

We now come to postulate (3) for gravity. This postulate relates General Relativity to Newtonian gravity, and Newtonian gravity is now known to be seriously flawed, and so *ipso facto,* by using this postulate, Einstein and Hilbert inadvertently developed a flawed theory of gravity. Newtonian gravity was based upon Kepler's Laws for the planetary motions within the solar system and uses the acceleration field $g$,

$$\nabla . g = -4\pi G \rho \tag{2}$$

where $G$ is Newton's universal gravitational constant, and $\rho$ is the density of matter. However equally valid mathematically is a velocity field formulation [2,3,7,8]

$$\frac{\partial}{\partial t}(\nabla . v) + \nabla .((v.\nabla)v) = -4\pi G \rho \tag{3}$$

with $g$ now given by the Euler `fluid' convective acceleration

$$g = \frac{\partial v}{\partial t} + (v.\nabla)v = \frac{dv}{dt} \tag{4}$$

External to a spherical mass $M$ a static velocity-field solution is

$$v(r) = -\sqrt{\frac{2GM}{r}}\hat{r} \tag{5}$$

which gives from (4) the usual inverse square law

$$g(r) = -\frac{GM}{r^2} \tag{6}$$

However (3) is not uniquely determined by Kepler's laws because

$$\frac{\partial}{\partial t}(\nabla.v) + \nabla.((v.\nabla)v) + C(v) = -4\pi G\rho \tag{7}$$

where

$$C(v) = \frac{\alpha}{8}((trD)^2 - tr(D^2)) \tag{8}$$

and

$$D_{ij} = \frac{1}{2}\left(\frac{\partial v_i}{\partial x_j} + \frac{\partial v_j}{\partial x_i}\right) \tag{9}$$

also has the same external solution (5), as $C(v) = 0$ for the flow in (5). So the presence of the $C(v)$ dynamics would not have manifested in the special case of planets in orbit about the massive central sun. Here α is a dimensionless constant - a new additional gravitational constant. However inside a spherical mass $C(v) \neq 0$ and using the Greenland ice-shelf bore hole $g$ anomaly data, Fig.3, we find [6] that $\alpha^{-1} = 139 \pm 5$, which gives the fine structure constant α = 1/137 to within experimental error.

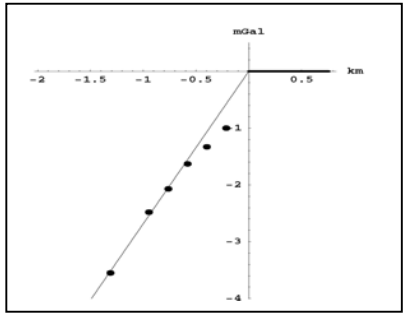

Fig.3 Gravity residuals from the Greenland bore hole anomaly data [17]. These are the differences in g between the measured g and that predicted by the Newtonian theory. According to (7) this difference only manifests within the earth, see [6], and so permits the value of α to be determined. The data shows that α is the fine structure constant, to within experimental error. This small value for α explains why the spiral galaxy rotation speed plots are so flat, and also explains the black holes masses at the centre of globular clusters.

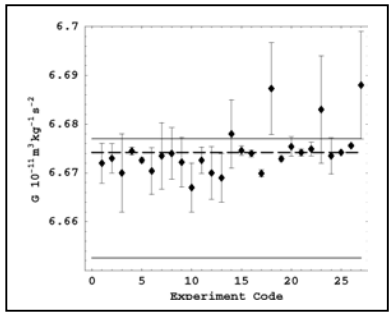

Fig.4 Results of precision measurements of G over the last sixty years in which the Newtonian theory of gravity was used to analyse the data. Shows systematic effect missing from the Newtonian theory, of fractional size ≈α/4. For this reason G is the least accurately known fundamental constant. The upper horizontal line shows the value of G from an ocean Airy measurement [18], while the dashed line shows the current CODATA value. The lower horizontal line shows the value of G after correcting for the `dark matter' effect. Results imply that Cavendish laboratory experiments can measure α.

From (7) and (8) we can introduce an additional effective `matter density'

$$\rho_{DM} = \frac{\alpha}{32\pi G}((trD)^2 - tr(D^2)) \tag{10}$$

as a phenomenological treatment of the new space dynamics within the velocity formulation of Newtonain gravity in (3). It is this spatial dynamics that has been misinterpreted as the `dark matter' effect. This `dark matter' dynamical effect also appears to explain the long-standing problems in measuring $G$ in Cavendish-type experiments, as shown in Fig.4.

Eqn.(7) has novel black hole solutions [6] where the in-flow is given by

$$v(r) = K\left(\frac{1}{r} + \frac{1}{R_S}\left(\frac{R_S}{r}\right)^{\frac{\alpha}{2}}\right)^{1/2} \tag{11}$$

where the key feature is the $\alpha$-dependent term in addition to the usual `Newtonian' in-flow in (5). For the in-flow in (11) the centripetal acceleration relation $v_o = \sqrt{rg(r)}$ for circular orbits gives orbital rotation speeds of the form

$$v(r) = \frac{K}{2}\left(\frac{1}{r} + \frac{\alpha}{2R_S}\left(\frac{R_S}{r}\right)^{\frac{\alpha}{2}}\right)^{1/2} \tag{12}$$

which is characterised by their almost flat asymptotic limit. This rotation curve explains the `dark matter' effect as seen in spiral galaxies, as shown in Fig.5.

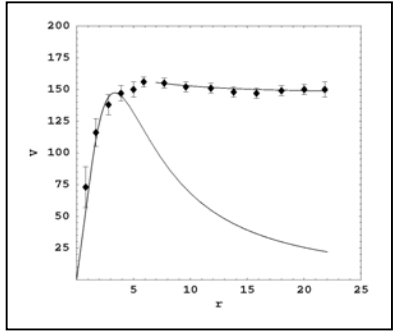

Fig.5 Data shows the non-Keplerian rotation-speed curve for the spiral galaxy NGC3198 in km/s plotted against radius in kpc/h. Complete curve is the rotation curve from the Newtonian theory or from General Realtivity for an exponential disk, which decreases asymptotically like $1/\sqrt{r}$. This descrepency is the origin of the `dark matter' story. The full curve shows the asymptotic form from (12), with the decrease determined by the small value of α. This asymptotic form is caused by the primordial black holes at the centres of spiral galaxies, and which play a critical role in their formation. The spiral structure is casued by the rapid in-fall to these primoridal black holes.

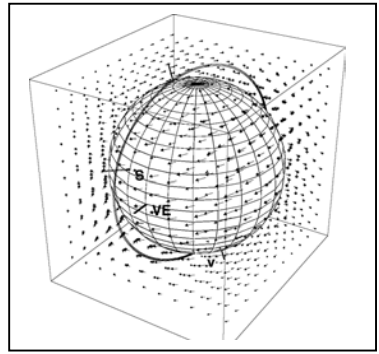

Fig.6 Shows the earth with absolute linear velocity V in the direction discovered by Miller [5] in 1925/26. This motion causes a vorticity, shown by the vector field lines. A much smaller vorticity is generated by the rotation of the earth, known as the Lenze-Thirring effect. In General Relativity only the earth-rotation induced vorticity is permitted, where it is known as a gravitomagnetic effect. Vorticity is a local rotation of space relative to more distant space. This rotation can be detected by observing the precession of a gyroscope, whose spin direction S remains fixed in the local space. VE is the vernal equinox. The Gravity Probe B satellite experiment [14,15] is designed to detect these precessions.

Eqn.(3) is only applicable to a zero vorticity flow. The vorticity is given by

$$\nabla \times (\nabla \times v) = \frac{8\pi G \rho}{c^2} v_R \qquad (13)$$

where $v_R$ is the absolute velocity of the matter relative to the local space. See [3,6,11] for the more general form of (7) that includes vorticity. Fig.6 shows the vorticity field $\nabla \times v$ induced by the earth's absolute linear motion. Eqn.(13) explains the Lense-Thirring `frame-dragging' effect in terms of vorticity in the flow field, but makes predictions very different from General Relativity. These conflicting predictions will soon be tested by the Gravity Probe B [14,15] gyroscope precession satellite experiment. However the smaller component of the frame-dragging effect caused by the earth's absolute rotation component of $v_R$ has been determined from the laser-ranged satellites LAGEOS(NASA) and LAGEOS2(NASA-ASI) [16] and the data implies the indicated coefficient on the RHS of (11) to ±10%. However that experiment cannot detect the larger component of the `frame-dragging' or vorticity induced by the absolute linear motion component of the earth as that effect is not cumulative, while the rotation induced component is cumulative. The GP-B gyroscope spin precessions caused by the earth's absolute motion are shown in Fig.7.

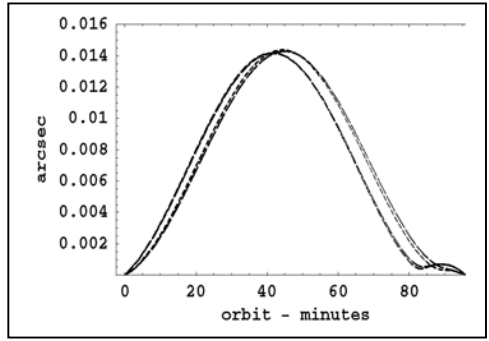

Fig.7 Shows predicted spin precession angle for the Gravity Probe B satellite experiment over one orbit, with the orbit shown in Fig.6, caused by the vorticity arising from the absolute linear motion of the earth [11,12]. Plots for four different months. This particular precession is not cumulative, compared to the precession from the earth-rotation induced vorticity component. The GP-B experiment is optimized to detect this cumulative spin precession. General Relativity has earth-rotation induced vorticity but not linear motion induced vorticity.

Both (3) and (7) have wavelike aspects to their time-dependent solutions, with the time-dependence being the rule rather than the exception [2,3,7,12]. Such wave behaviour has been detected in most absolute motion experiments, as seen in the DeWitte data in Fig.2. For (3) these waves do not produce a gravitational force effect via (4), but with the inclusion of the `dark matter' spatial dynamics in (7) such waves do produce a gravitational force effect, and there are various experimental `anomalies` which are probably manifestations of this effect. As well such waves affect the vorticity from (13), and in principle could be detected by the GP-B experiment. General Relativity predicts a very different kind of gravitational wave, but these have never been seen, despite extensive searches. The new theory of gravity implies that these waves do not exist.

The trajectory of test particles in the differentially flowing space are determined by extremising the proper time

$$\tau[r_o] = \int dt \left(1 - \frac{v_R^2}{c^2}\right)^{1/2} \tag{14}$$

where $v_R = v - v_o$, with $v_o$ the velocity of the object relative to an observer frame of reference, which gives from (14) an acceleration independent of the test mass, in accord with the equivalence principle,

$$\frac{dv_o}{dt} = \left(\frac{\partial v}{\partial t} + (v.\nabla)v\right) + (\nabla \times v) \times v_R - \frac{v_R}{1 - \frac{v_R^2}{c^2}} \frac{1}{2} \frac{d}{dt}\left(\frac{v_R^2}{c^2}\right) \tag{15}$$

where the 1st term is the Euler `fluid' convective acceleration in (4), the 2nd term is the vorticity induced Helmholtz acceleration, and the last is a relativistic effect leading to the `geodesic' effects. It is significant that the time-dilation effect in (14) leads to well known `fluid` accelerations, revealing a close link between the spatial flow phenomena and relativistic effects.

## General Relativity Flow-Metrics

We saw that Newtonian gravity failed because it was expressed in the limited formalism of the gravitational acceleration field $g$. As soon as we introduce the velocity field formalism together with its `dark matter' generalisation we see that numerous gravitational anomalies are explained [6]. General Relativity was constructed to agree with Newtonian gravity, and so it is flawed by this connection. So it is interesting to understand why General Relativity (GR) is supposed to have passed key observational and experimental tests. GR uses the Einstein tensor

$$G_{\mu\nu} \equiv R_{\mu\nu} - \frac{1}{2} R g_{\mu\nu} = \frac{8\pi G}{c^2} T_{\mu\nu} \tag{16}$$

In this formalism the trajectories of test objects are also determined by extremising (14) which, after a general change of coordinates, gives the acceleration in (15) in terms of the usual affine connection.

$$\Gamma^\lambda_{\mu\nu} \frac{dx^\mu}{d\tau} \frac{dx^\nu}{d\tau} + \frac{d^2 x^\lambda}{d\tau^2} = 0 \tag{17}$$

In the case of a spherically symmetric mass $M$ the well known solution of (16) outside of that mass is the external-Schwarzschild metric

$$d\tau^2 = \left(1 - \frac{2GM}{c^2 r}\right) dt^2 - \frac{r^2}{c^2}\left(d\theta^2 + \sin^2(\theta) d\varphi^2\right) - \frac{dr^2}{c^2\left(1 - \frac{2GM}{c^2 r}\right)} \tag{18}$$

This solution is the basis of various experimental checks of General Relativity in which the spherically symmetric mass is either the sun or the earth. The four tests are: the gravitational redshift, the bending of light, the precession of the perihelion of Mercury, and the time delay of radar signals. However the solution (18) is in fact completely equivalent to the in-flow interpretation of Newtonian gravity. Making the change of variables $t \to t'$ and $r \to r' = r$ with

$$t' = t + \frac{2}{c}\sqrt{\frac{2GMr}{c^2}} - \frac{4GM}{c^2} \tanh^{-1} \sqrt{\frac{2GM}{c^2 r}} \tag{19}$$

the Schwarzschild solution (18) takes the form

$$d\tau^2 = dt'^2 - \frac{1}{c^2}\left(dr' + \sqrt{\frac{2GM}{r'}}dt'\right)^2 - \frac{r'^2}{c^2}\left(d\theta'^2 + \sin^2(\theta')d\varphi'^2\right) \tag{20}$$

which is exactly the differential form of (14) for the velocity field given by the Newtonian form in (5). This choice of coordinates corresponds to a particular frame of reference in which the test object has velocity $v_R$ relative to local space. This result shows that the Schwarzschild metric in GR is completely equivalent to Newton's inverse square law: GR in this case is nothing more than Newtonian gravity in disguise. So the so-called `tests' of GR were nothing more than a test of the `geodesic' equation (14), where most simply this is seen to determine the motion of an object relative to an observable and observed absolute local frame of reference. These tests were merely confirming the in-flow formalism, and have nothing to do with a Schwarzschild spacetime ontology.

Since GR has only been directly tested using the metric in (18) or (20), it is interesting to ask what is the particular form that (16) then takes. To that end we substitute a special class of flow-metrics, involving an arbitrary time-dependent velocity flow-field, into (16)

$$d\tau^2 = g_{\mu\nu}dx^\mu dx^\nu = dt^2 - \frac{1}{c^2}\left(dr - v(r,t)dt\right)^2 \tag{21}$$

The various components of the Einstein tensor are then found to be

$$G_{00} = \sum v_i \overline{G}_{ij} v_j - c^2 \sum \overline{G}_{0j} v_j - c^2 \sum v_i \overline{G}_{i0} + c^2 \overline{G}_{00}$$
$$G_{i0} = -\sum \overline{G}_{ij} v_j + c^2 \overline{G}_{io}$$
$$G_{ij} = \overline{G}_{ij} \tag{22}$$

where the $\overline{G}_{\mu\nu}$ are given by

$$\overline{G}_{00} = \frac{1}{2}((trD)^2 - tr(D^2))$$
$$\overline{G}_{i0} = \overline{G}_{0i} = -\frac{1}{2}\left(\nabla \times (\nabla \times v)\right)_i$$
$$\overline{G}_{ij} = \frac{d}{dt}\left(D_{ij} - \delta_{ij}trD\right) + \left(D_{ij} - \frac{1}{2}\delta_{ij}trD\right)trD - \frac{1}{2}\delta_{ij}tr(D^2) - (D\Omega - \Omega D)_{ij} \tag{23}$$

and where $\Omega_{ij} = \frac{1}{2}\left(\frac{\partial v_i}{\partial x_j} - \frac{\partial v_j}{\partial x_i}\right)$ is the tensor form of the flow vorticity. In vacuum, with $T_{\mu\nu} = 0$, we find that $G_{\mu\nu} = 0$ implies that $\overline{G}_{\mu\nu} = 0$. This system of GR equations then demands that

$$\rho_{DM} = \frac{\alpha}{32\pi G}((trD)^2 - tr(D^2)) = 0 \tag{24}$$

This simply corresponds to the fact that GR does not permit the `dark matter' effect, and this happens because GR was forced to agree with Newtonian gravity, in the appropriate limits, and that theory also has no such effect. As well in GR the energy-momentum tensor $T_{\mu\nu}$ is not permitted to make any reference to absolute linear motion of the matter; only the relative motion of matter or ab-solute rotational motion is permitted. It is very significant to note that the above exposition of the GR formalism for the metrics in (21) is exact. Taking the trace of $\overline{G}_{ij}$ in (22) we obtain, also exactly, and in the case of zero vorticity and outside of matter, that

$$\frac{\partial}{\partial t}(\nabla.v) + \nabla.((v.\nabla)v) = 0 \qquad (25)$$

which is exactly the `velocity field' formulation in (7) of Newtonian gravity outside of matter. This should have been expected as it corresponds to the previous observation that the `Newtonian in-flow' velocity field is exactly equivalent to the external-Schwarzschild metric. There is in fact only one definitive confirmation of the GR formalism apart from the misleading external-Schwarzschild metric cases, namely the observed decay of the binary pulsar orbital motions, for only in this case is the metric non-Schwarzschild, and therefore non-Newtonian. However the new theory of gravity also leads to the decay of orbits, and on the grounds of dimensional analysis we would expect comparable predictions. It is also usually argued that the Global Positioning System (GPS) demonstrated the efficacy of General Relativity. However as shown in [13] the new flow formalism of gravity also explains this system, and indeed gives a physical insight into the processes involved. In particular the relativistic speed and red-shift effects now acquire a unified explanation.

## Discussion and Conclusions

The experimental evidence from at least seven observations of absolute linear motion, some using Michelson interferometers and some coaxial cable experiments, all showed that absolute linear motion is detectable, and indeed has been so ever since the 1887 Michelson-Morley experiment. Even Michelson and Morley reported a speed of 8km/s using the Newtonian theory for the instrument, but which becomes $\geq v_p = 300$ km/s when the Fitzgerald-Lorentz dynamical contraction effect and the refractive index effect are both taken into account. It then follows that vacuum interferometer experiments will fail to detect that absolute motion, as is the case. We also understand that the various relativistic effects are caused by the absolute motion of systems through space, an idea that goes back to Lorentz. Elsewhere [2,3,7] we have shown that both the Galilean and Lorentz transformations have a role in describing mappings of data between observers in relative motion, but that they apply to different forms of the data. So absolute motion is a necessary part of the explanation of relativistic effects, and indeed the Lorentz transformations and symmetry are consistent with absolute motion, contrary to current beliefs. On the contrary the Einstein postulates and their apparent link to these relativistic effects have always been understood to imply that absolute motion is incompatible with these relativistic effects. It was then always erroneously argued that the various observations of absolute motion over more than 100 years must have been flawed, since the relativistic effects had been confirmed in numerous experiments.

So the Einstein postulates have had an enormously negative influence on the development of physics, and it could be argued that they have resulted essentially in a 100-year period of stagnation of physics, despite many other exciting and valid developments, but even these will require a review of their deeper foundations, particularly in the case of electromagnetism.

A major effect of the Einstein postulates was the development of a relativistic theory of gravity that was constrained to agree with Newtonian gravity in the non-relativistic limit. Evidence that Newtonian gravity was flawed has been growing for over 50 years, as evidenced by the numerous so-called `gravitational anomalies', namely experimental observations of gravitational effects incompatible with Newtonian gravity. The most well known of these is the `dark matter' effect, namely that spiral galaxies appear to require at least 10x the observed matter content in order to explain the high rotation speeds of stars and gas clouds in the outer regions. We now see that this effect is not caused by any form of matter, but rather by a non-Newtonian aspect to gravity. As well the Greenland bore hole g anomaly data has revealed that the dimensionless constant that determines the magnitude of this spatial dynamics is non other than the fine structure constant. The detection of absolute motion implies that space has some structure for it is motion through that structure which is known as `absolute motion', and which is causing relativistic effects. This means that the phenomena of gravity are described by two gravitational constant, $G$ and $\alpha$, and it is the small size of $\alpha$ that determines the asymptotic form of the orbital rotation speeds in spiral galaxies. As well it is $\alpha$ that determines the magnitude of the black hole masses at the centres of globular clusters, and the data from M15 and G1 confirms that $M_{BH} = \alpha M_{GC}/2$, in agreement with (7), see [6].

The detection of absolute motion and the failure of Newtonian gravity together imply that General Relativity is not a valid theory of gravity; and that it is necessary to develop a new theory. This has now been achieved, and the essential task of checking that theory against experiment and observation has now explained all the known effects that GR was supposed to have explained, but most significantly, has also explained the numerous `anomalies' where GR was in manifest disagreement with the experimental or observational data. In particular a component of the flow past the earth towards the sun has been extracted from the analysis of the yearly variations of the Miller data [2,3,4].

The putative successes of the Einstein postulates lead to the Minkowski-Einstein spacetime ontology that has dominated the mindset of physicists for 100 years. Spacetime was mandated by the misunderstanding that absolute motion had not been observed, and indeed that it was incompatible with the established relativistic effects. Of course it was always possible to have chosen one foliation of the spacetime construct as the actual one separating the geometrical model of time from the geometrical model of space, but that never happened, and that possibility became one of the banned concepts of physics.

We are now in the position of understanding that space is a different phenomenon from time, that they are not necessarily fused into some spacetime amalgam, and that the spacetime ontolgy has been one of the greatest blunders in physics. This must not be misunderstood to imply that the numerous uses of a *mathematical* spacetime, particularly in Quantum Field Theory, were invalid. What is invalid is the assertion that such a spacetime is a physical entity.

We may now ask, for the first time in essentially 100 years, about the nature of space. It apparently has `structure' as evidenced by the fact that motion through it is detectable by various experimental techniques, and that its self-interaction is determined in part by the fine structure constant. As argued elsewhere [2,3,7] one interpretation is that space is a quantum system undergoing classicalisation, and at a deep substratum level has the characteristics of a quantum foam. This quantum foam is in differential motion, and the inhomogeneities and time dependencies of this motion cause accelerations which we know of as gravity. This motion is not motion of something through a geometrical space, but an ongoing restructuring of that quantum foam. One theory for this quantum foam arises in an information-theoretic formulation of reality known as *Process Physics*, and one implication of that is that the quantum-foam system undergoes exponential growth, once the size of the quantum-foam system dominates over the matter part of the universe. This effect then appears to explain what is known as the `dark energy' effect, although of course it is not an energy at all, just as `dark matter' is not a form of matter. As well within this *Process Physics* we see a possible explanation for quantum matter, namely as topological defects embedded in the spatial quantum foam. That works gives the first insights into an explanation for the necessity of quantum behaviour and classicalisation.

This quantum-foam spatial system invites comparison with the much older concept of the 'aether', but it differs in that the aether was usually considered to be some form of matter residing within a geometrical space, which is not the case here with the quantum foam theory of space; for here the geometrical description of space is merely a coarse grained description. Nevertheless it would be uncharitable not to acknowledge that the quantum-foam system is a modern version and indeed a return to the aether concept, albeit a banned concept.

Physics is a science. This means that it must be based on (i) experiments that test its theories, and (ii) that its theories and reports of the analyses of experimental outcomes must be freely reported to the physics community. Regrettably, and much to its detriment, this has ceased to be the case for physics. Physics has been in an era of extreme censorship for a considerable time; Miller was attacked for his major discovery of absolute linear motion in the 1920's, while DeWitte was never permitted to report the data from his beautiful 1991 coaxial cable experiments. Amazingly these experimenters were unknown to each other, yet their data was is in perfect agreement, for by different techniques they were detecting the same phenomenon, namely the absolute linear motion of the earth through space. All discussions of the experimental detections of absolute motion over the last 100 years are now banned from the mainstream physics publications. But using modern vacuum resonant cavity interferometer technology, and with a gas placed in the cavities, these devices could be used to perform superb experimental detections of absolute motion. As well the Miller and DeWiite data shows the presence of a wave phenomenon different to the waves argued to arise within GR theory, but which have not been detected, de-spite enormous costly efforts. It is now time to separate the genuine relativistic effects and their numerous manifestations from the flawed Einstein postulates, and to finally realise that they are caused by absolute motion of systems through a complex quantum system, which we know of as space. As for General Relativity it turns out to have been a major blunder. Nevertheless there is much evidence that a new theory of gravity has emerged, and this is to be exposed to critical analysis, and experimental and observational study.